\documentclass[aps,pre,twocolumn,superscriptaddress,floatfix,10pt]{revtex4-1}
\usepackage{graphicx}
\usepackage{dcolumn}
\usepackage{bm}
\usepackage{latexsym}
\usepackage{amsmath}
\usepackage{amssymb}
\usepackage{color}

\begin{document}


\title{A biologically inspired two-species exclusion model:\\ 
effects of RNA polymerase motor traffic on simultaneous DNA replication}
 \author{Soumendu Ghosh}
\affiliation{Department of Physics, Indian Institute of Technology
  Kanpur, 208016, India}
 \author{Bhavya Mishra}
\affiliation{Department of Physics, Indian Institute of Technology
  Kanpur, 208016, India}

\author{Shubhadeep Patra}
\affiliation{ISERC, Visva-Bharati, Shantiniketan 731235, India}
 \author{Andreas Schadschneider}
\affiliation{Institute for Theoretical Physics, University of Cologne, 
K\"oln, Germany}
\author{Debashish Chowdhury{\footnote{Corresponding author; e-mail: debch@iitk.ac.in}}}
\affiliation{Department of Physics, Indian Institute of Technology
  Kanpur, 208016, India}

\begin{abstract}

  We introduce a two-species exclusion model to describe the key
  features of the conflict between the RNA polymerase (RNAP) motor
  traffic, engaged in the transcription of a segment of DNA,
  concomitant with the progress of two DNA replication forks on the
  same DNA segment. One of the species of particles ($P$) represents
  RNAP motors while the other ($R$) represents replication
  forks. Motivated by the biological phenomena that this model is
  intended to capture, a maximum of only two $R$ particles are allowed
  to enter the lattice from two opposite ends whereas the unrestricted
  number of $P$ particles constitute a totally asymmetric simple
  exclusion process (TASEP) in a segment in the middle of the
  lattice. Consequently, the lattice consists of three segments; the
  encounters of the $P$ particles with the $R$ particles are confined
  within the middle segment (segment $2$) whereas only the $R$
  particles can occupy the sites in the segments $1$ and $3$. The
  model captures three distinct pathways for resolving the
  co-directional as well as head-collision between the $P$ and $R$
  particles. Using Monte Carlo simulations and heuristic analytical
  arguments that combine exact results for the TASEP with mean-field
  approximations, we predict the possible outcomes of the conflict
  between the traffic of RNAP motors ($P$ particles engaged in
  transcription) and the replication forks ($R$ particles). The
  outcomes, of course, depend on the dynamical phase of the TASEP of
  $P$ particles. In principle, the model can be adapted to the
  experimental conditions to account for the data quantitatively.

\end{abstract}

\maketitle

\section{Introduction}

The totally asymmetric simple exclusion process (TASEP)
\cite{derrida98,schutz00,Schadschneider10,mallick15} was originally
introduced as a simplified model describing the kinetics of protein
synthesis \cite{MacDonald68,macdonald69}. Since then it has found many
more applications to biological systems, especially to situations
\cite{basu07,gccr09,zia11,lin11,greulich12,chowdhury08,oriola15,sugden07,evans11,chai09,ebbinghaus09,ebbinghaus10,muhuri10,neri11,neri13a,neri13b,curatolo16,klein16,parmeggiani04,graf17},
where the kinetics is dominated by the traffic-like collective motion
of molecular motors (for reviews, see
\cite{chowdhury05,chou11,chowdhury13,rolland15,kolomeisky15}).
Genetic message encoded chemically in the sequence of the monomeric
subunits of DNA is “transcribed” into an RNA molecule by a molecular
motor called RNA polymerase (RNAP).  In each of its step on the DNA
track a RNAP motor elongates the nascent RNA molecule by a single
subunit using the same DNA strand as the corresponding template
\cite{kornbergbook}.  TASEP-based models have also been developed for
the traffic-like collective movements of RNAP motors on the same
segment of DNA while each RNAP synthesizes a distinct copy of the same
RNA \cite{tripathi08,klumpp08,klumpp11,sahoo11,ohta11,wang14}.

A segment of DNA can undergo multiple rounds of transcription during
the lifetime of a cell.  In contrast, each DNA molecule is replicated
once, and only once, just before the cell divides into two daughter
cells \cite{kornbergbook}. A molecular machine called DNA polymerase
(DNAP) is a key component of a replisome which is a multi-machine
macromolecular complex that replicates DNA. As the replisomes unzip a
duplex DNA and replicate the two exposed strands, Y-shaped junctions
called replication forks, are formed. The progress of replication can
be described in terms of the movement of two replication forks;
replication of a segment of DNA is completed when two replication
forks, approaching each other from opposite ends of the segment,
collide head-on \cite{kornbergbook}.  Theoretical models for the
``nucleation'' of replication competent replication forks and growth
of the replicated domains of the DNA have been reported in the past
\cite{jun05a,jun05b,bechhoefer07,yang08,baker12,retkute11,retkute12}
(see also \cite{yang09,hyrien10} for reviews).

Interestingly, transcription and replication can occur simultaneously
on the same segment of DNA. However, typically, at a time only one of
the two DNA strands of the DNA undergoes transcription by a traffic of
RNAPs while both the strands are simultaneously replicated by distinct
replisomes.  Obviously, head-on collisions between a replication fork
and RNAP motors is possible. Moreover, since the rate of replication
is 10-20 times faster than that of transcription, a replication fork
can catch up with a RNAP from behind thereby causing co-directional
collision. Both types of collisions can have disastrous consequences
\cite{kim12}, unless the transcription-replication conflict is
resolved sufficiently rapidly to ensure maintenance of genomic
stability.  Nature has adopted multiple mechanisms of resolution of
such conflict \cite{merrikh12,helmrich13,muse16,pomerantz10}.
However, to our knowledge, no quantitative theoretical model of
transcription-replication conflict and their resolution has been
reported so far.

Here we propose a TASEP-based minimal model that captures the
essential aspects of RNAP traffic on a segment of DNA concomitant with
the progress of DNA replication forks from the two ends of the same
DNA segment. The kinetics of the model incorporates all the known
natural mechanisms of resolution of conflicts between DNA replication
and transcription. This formulation, as explained in the next section,
leads to a two-species exclusion process on a 3-segment lattice in one
dimension which also includes ``Langmuir kinetics'', i.e. attachment and
detachment of particles in the bulk \cite{parmeggiani03}.  One of
the two species of particles represents RNAP motors all of which move
co-directionally, i.e., say, from left to right. In contrast, only two
particles of the second species, each representing a DNA replication
fork, approach each other from opposite ends of the same track, i.e.,
one from the left and the other from the right.  Because of the
decrease of the separation between the two replication forks with the
passage of time, the spatial region of conflict between the two
species of particles also keeps shrinking. Thus, the model of the
two-species exclusion process developed here is highly non-trivial. 

By a combination of analytical arguments and computer simulations, we
investigate the effect of the two processes, i.e., transcription and
replication, on each other. 
More specifically, we indicate (a) the trends of variation 
of the mean time for completion of replication and (ii) the statistics of the 
successful and unsuccessful replication events, in the different phases of 
the RNAP traffic \cite{schutz00,Schadschneider10,mallick15}.

\section{Model}
\begin{figure}[t]
    \begin{center}
        \includegraphics[angle=0,width=1.0\columnwidth]{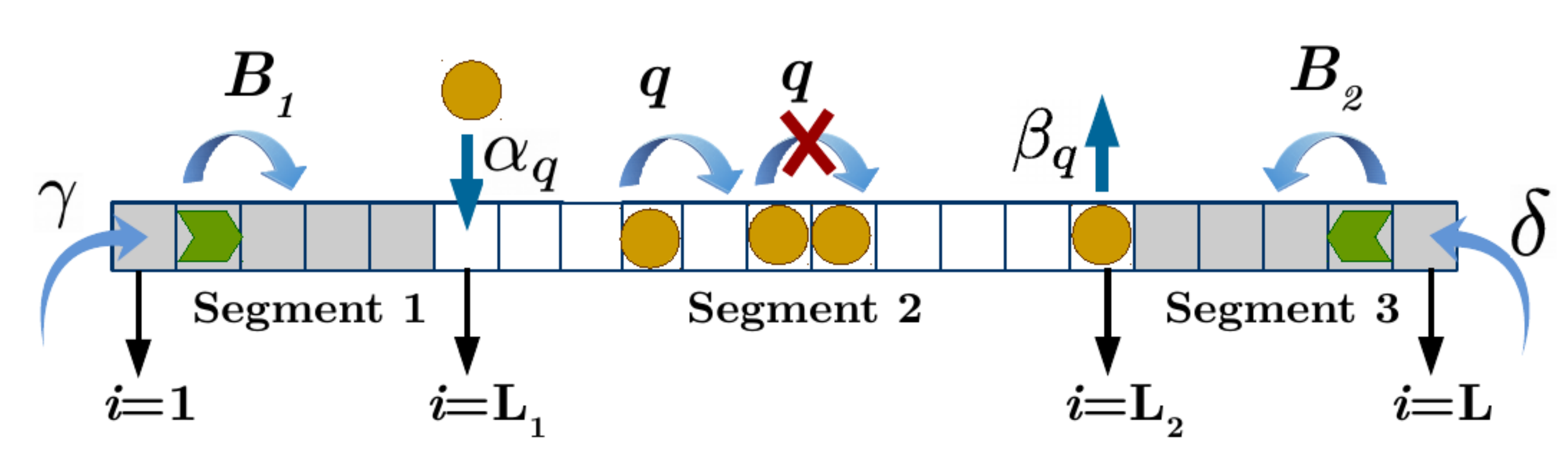}\\[0.02 cm]
    \end{center}
    \caption{A schematic diagram of the model. The whole lattice is
      divided into three segments $\{1,2,\ldots,L_1-1\}$,
      $\{L_1,\ldots,L_2\}$ and $\{L_2,\ldots,L\}$. 
      A $R$ particle (green arrow) can enter, either from the first
      site of segment $ 1 $ (i.e. $ i=1 $) with the probability
      $ \gamma $ per unit time or from the last site of segment $ 3 $
      (i.e. $ i=L $), with the probability $ \delta $ per unit time. A
      $R$ particle that enters through $i=1$ is allowed to hop from
      left to right (i.e. $ i \rightarrow i+1 $) with rate $ B_1
      $, if the target site is empty.
      But, if a $R$ particle enters through $i=L$ it is allowed to
      hop only from right to left (i.e. $ i \rightarrow i-1$), with
      rate $B_2$.  Both the $R$ particles continue their motion until
      they meet each other, at a pair of nearest neighbour
      sites. However, inside segment $ 2 $, a new $P$ particle (yellow 
circle) can
      attach only at $ i=L_1 $, with rate $ \alpha_q $, only if this
      site is empty.  Once attached, a $P$ particle can hop forward
      only from left to right (i.e.  $ i \rightarrow i+1 $) by a
      single site in each step, with rate $ q $, provided the target
      site is empty. Normally, a $P$ particle would continue its
      hopping till it reaches the the site $L_2$ from where it
      detaches with rate $ \beta_q $. Thus, the lattice sites in the
      segments $ 1 $ and $ 3 $ can be occupied by only the $R$
      particles, whereas a mixed population of $R$ and $P$ particles
      can exist in the segment $ 2 $.}
    \label{fig-model-tasep}
\end{figure}

The schematic diagram of the model is shown in
Fig.~\ref{fig-model-tasep}. For simplicity, motion of both species of
particles are assumed to occur along a single common track represented
by a one dimensional lattice of total length $L$, where, $L$ is the
total number of equispaced sites on the lattice.  The lattice consists
of three segments: sites $i=1$ to $i=L_{1}-1$ (segment $1$), $i=L_{1}$
to $i=L_{2}$ (segment $2$), and site $i=L_{2}+1$ to $i=L$ (segment
$3$).

One of the two species of particles, labelled by $P$, represent the
RNAP motors; all the $P$ particles can move, by convention, only from
left to right, i.e., from $i$ to $i+1$. There is no restriction on the
number of $P$ particles that can populate the lattice, except the
limits arising naturally from the rates of entry, exit and forward
hopping that are described below. In contrast, not more than two
particles of the second species, labelled $R$ and representing the
replication forks, can ever enter the lattice irrespective of the
kinetic rates, i.e., probabilities per unit time of the various
kinetic processes that are described below. One of the $R$ particles,
denoted by $R_{\ell}$ moves from left to right ($i$ to $i+1$) whereas
the other, denoted by $R_{r}$ moves from right to left ($i+1$ to $i$)
on the lattice.

The $R_{\ell}$ particle can enter the lattice only at $i=1$ with the
probability $\gamma$ per unit time.  Similarly, the particle $R_{r}$
can enter the lattice only at $i=L$ with the probability $\delta$ per
unit time. After entry, the particles $R_{\ell}$ and $R_{r}$ can hop to
the next site in their respective pre-determined directions of motion
with the rates $B_{1}$ and $B_{2}$, respectively (see
Fig.~\ref{fig-model-tasep}). Both these particles can continue hopping,
obeying the exclusion principles and rules of resolution of encounter
with $P$ particles as described below, till they encounter each other
head-on at two nearest-neighbor sites on the lattice indicating
completion of replication.

Unlike the $R$ particles, all the $P$ particles can enter the lattice
only at the site $i=L_{1}$ with the attachment rate (i.e., probability
per unit time) $\alpha_{q}$ provided that site is not already occupied
by any other particle of either species. Once entered, a $P$ particle
can hop forward to the next site with the rate $q$ if, and only if,
the target site is not already occupied by any other $P$ or $R$
particle.  A $P$ particle can continue forward hopping, obeying the
exclusion principle and the rules of resolution of encounter with $R$
particles, till it reaches the site $i=L_{2}$ from where it can exit
with the rate $\beta_{q}$.

Thus, the division of lattice into three segments is based on the
scenario that the lattice sites in the segments $1$ and $3$ can be
occupied exclusively by only the $R$ particles whereas the sites in
middle region (i.e. segment $2$) can get populated by both the $P$ and
$R$ particles. However, in the segment $2$ the $P$ particles encounter
the $R_{\ell}$ particle co-directionally and $R_{r}$ particle head-on.
The final encounter between the two $R$ particles, when they meet each
other at two nearest-neighbour sites, is head-on.

Next we described the kinetics of both types of particles in the
segment $2$ which capture the mutual exclusion of the RNAP motors as
well as the rules of resolution of the conflicts between transcription
and replication. Mutual exclusion is captured by the simple rule that
no site can be occupied simultaneously by more than one particle
irrespective of the species to which it belongs. The three possible
outcomes of the encounter between a $P$ particle at the site $i$ and a
$R_{\ell}$ particle at site $i-1$ or a $R_{r}$
particle at the site $i+1$ are as follows:\\
(a) The $R$ particle can bypass the $P$ particle with the rates
$ p_{\text{co}} $ and $ p_{\text{contra}} $, in the cases of co-directional and
contra-directional encounter respectively, without dislodging the
latter from the lattice and, therefore, both the particles can
continue hopping in their respective natural direction of
movement after the encounter.\\
(b) The $R$ particle can knock the $P$ particle out of the track, with
the rate $D$ irrespective of the direction (co- or contra-directional)
encounter and it resumes its hopping after the $P$ particle is swept
out of its way thereby aborting the transcription by that $P$ particle 
prematurely. \\
(c) Upon encounter with $P$ particles, a $R$ particle does not
necessarily always win. In such situations, occasionally, the $R$
particle detaches from the lattice with a probability $C$ per unit
time irrespective of the direction of encounter; this scenario
captures the possible collapse of the replication fork that can causes
genome instability. Once the replication fork collapses, the
victorious $P$ particle(s) resume their onward journey on the lattice.

\begin{figure}[h]
    \begin{center}
        \includegraphics[angle=0,width=0.7\columnwidth]{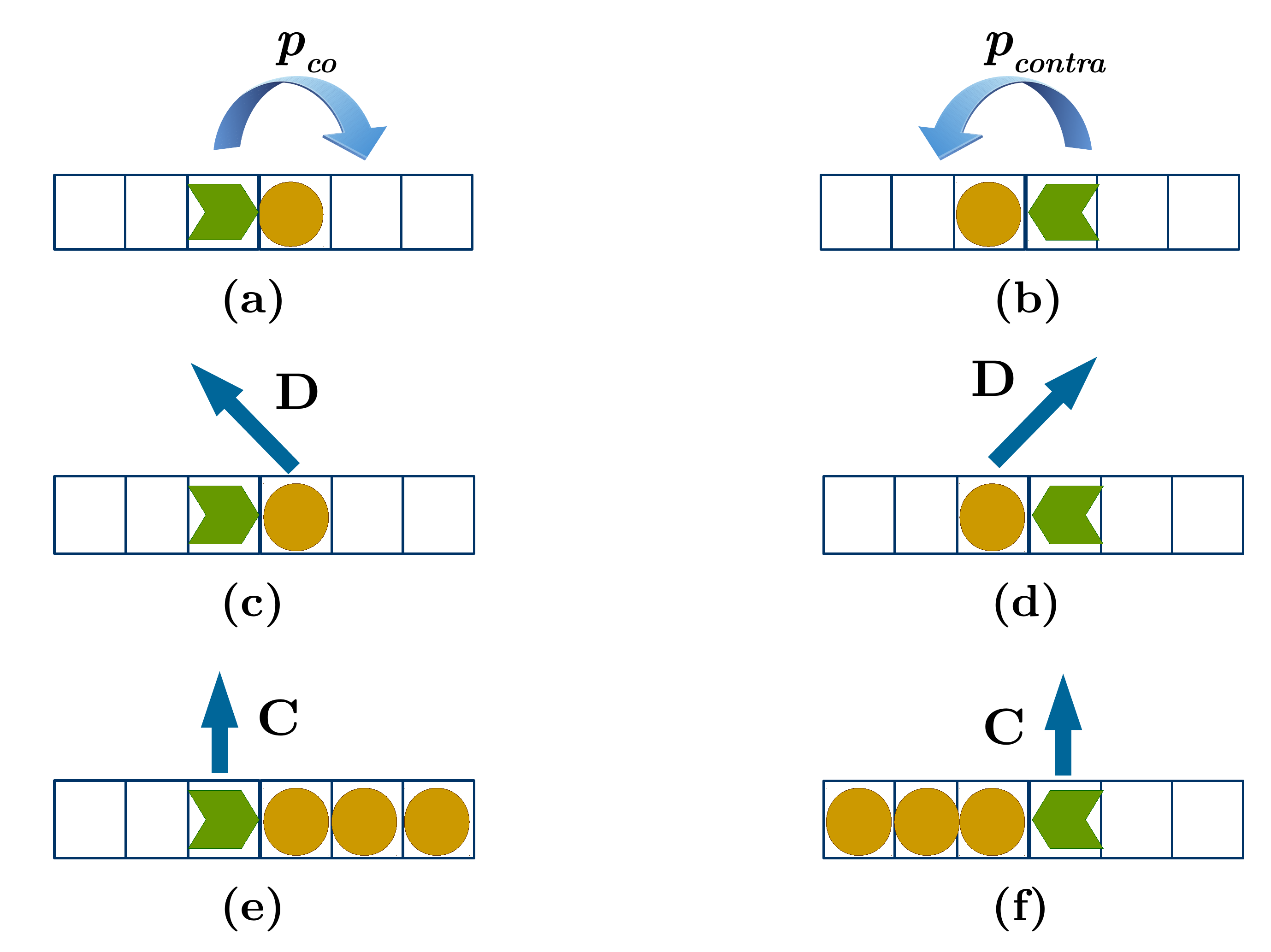}\\[0.02 cm]
    \end{center}
    \caption{Schematic representation of interference between the $R$
      particles and the $P$ particles. In (a) and (b) $R$ particle can
      pass the $P$ particle, with rates $ p_{\text{co}} $ and $ p_{\text{contra}}$.
      In (c) and (d) the $R$ particle can knock the $P$ particle out
      of the track, with rate $ D $. In (e) and (f) $P$ particle can
      block the progress of the $R$ particle thereby causing its
      eventual collapse, with rate $ C$. }
    \label{fig-model-bypass}
\end{figure}

If the $R$ particles, entered from the sites $i=1$ and $i=L$,
eventually meet each other on a pair of nearest-neighbour sites of the
lattice, thereby indicating completion of the replication of the
entire stretch of DNA from $i=1$ to $i=L$, we identify it as a
successful event of type 1 (from now onwards we referred to it to as
sr1). On the other hand, if one of the $R$ particle stalls or
collapses at any site in between $L_{1}$ and $L_{2}$ while the other
continues hopping until it reaches a nearest neighbor of that
particular site of stall or collapse, it also indicates successful
completion of replication and, therefore, identified as a successful
event of type 2 (from now onwards, referred to as sr2). But if both
the $R$ particles are stalled (i.e., replication fork collapsed)
before completely covering the entire lattice together by hopping
between the sites $1$ and $L$, then the process is identified as
unsuccessful event (usr). Dividing the sum total of the times taken by
all the sr1 and sr2 events by the total number of all such events we
obtain the mean hopping time of a $R$ particle ($\tau$), which is the
mean time required for successful completion of replication of the DNA
of length $L$ (in the units of ``base pairs'').

\section{Results}

Although our model captures just a few key aspects of the biological
processes involved in the transcription-replication conflict, the
proposed model is already too complex to allow a rigorous analytical
treatment. We therefore rely mainly on Monte Carlo (MC)
simulations. However, in certain limiting situations, the computer
simulations are complemented by an approximate analytical theory, that
draws heavily on the known exact results for TASEP with single species
of particles.  The transparent arguments of the analytical derivations
provide some insight into the underlying physical processes. However,
the analytical derivation is based primarily on heuristic arguments
some steps of which are essentially equivalent to mean-field
approximations. Therefore, the accuracy of our heuristic analytical
arguments have been checked by comparison with the corresponding data
obtained from the MC simulations.

In the MC simulations we adopted random sequential updating to
investigate the effects of traffic of $P$ particles on the kinetics of
$R$ particles, i.e, the effects of ongoing transcription on
replication. The data collected during the simulations are averaged
over 10000 realizations each starting from a fresh initial
configuration. We convert the rates into probabilities by using the
conversion formula $p_k=k~dt$ where, $k$ is an arbitrary rate constant
and $dt$ is an infinitesimally small time interval; the typical
numerical value of $dt$ used in our simulations is $ dt =
0.001~s$.
Unless stated explicitly otherwise, the numerical values of the
relevant parameters used in the simulations are $L=2000$, $L_{1}=500$,
$L_{2}=1500$, $B_{1}=B_{2}=300 ~s^{-1}$, $\beta_{q} = 1000 ~s^{-1}$
and $q=30 ~s^{-1}$.

\subsection{Effects of steady traffic of RNA polymersases on replication time}

The time needed for a successful completion of replication (now
onwards, referred to as ``replication time'') is identified as the
time taken by the two $R$ particles to meet head-on, starting from
their simultaneous entry into the lattice through $i=1$ and $i=L$. In
order to measure the replication time in the steady traffic of $P$
particles in the MC simulations, we first switch on the entry of only
the $P$ particles (i.e., transcription) through $i=L_{1}$.  The two
$R$ particles are allowed to enter simultaneously, through $i=L_{1}$
and $i=L_{2}$ only after the flux of the $P$ attains its constant
value in the non-equilibrium steady-state of the TASEP. Once the $R$
particles enter the segment $2$ and start encountering the $P$
particles, the rate of replication begins to get affected adversely.

We first present the derivation of the analytical results before
comparing with the corresponding data obtained from MC simulation.
Suppose $n$ denotes the number of particles in the segment $2$ of the
lattice. In the limit $ n\gg 1 $, the effects of a single $R$ particle
on the flow of the $P$ particles is expected to be negligibly small so
that the movement of the $P$ particles can be approximated well by a
purely single-species TASEP in the segment $2$. Under this assumption,
the flux $J_P$ of the $P$ particles corresponding to the number
density $\rho$ inside segment $2$ is given by the standard formula
(see, for example, \cite{chowdhury00})
\begin{equation}
 J_P= q \rho (1-\rho). 
 \label{flux_1}
\end{equation}
Since, because of the open boundaries, $n$ fluctuates with time even in
the steady state, the number density $\rho = n/(L_2-L_1)$, also
fluctuates with time. The effective velocity of the $P$ particles
corresponding to the flux (\ref{flux_1}) in segment $2$ is given by,
\begin{equation}
v_P=\dfrac{J_P}{\rho}=q(1-\rho).
\label{eff_vel_1}
\end{equation}

First we explore the parameter regime where $\beta$ is so large that
at sufficiently low values of $\alpha$ the $P$ particles would be in
the low-density (LD) phase of the TASEP (in the ``initiation''-limited
regime in the terminology of transcription). With the increase of
$\alpha$ the system would make a transition to the maximal current
(MC) phase of TASEP (``elongation-limited'' regime of transcription).
For the analytical derivation, we assume the following simplified situations:\\
(a) None of the $R$ particles collapse (i.e. $ C=0 $) upon encounter
with
$P$ particles,\\
(b) None of the $R$ particles can detach from the lattice prematurely
(i.e. $ D=0 $), \\
(c) In the absence of any hindrance, the rate of replication by both
the forks are identical, i.e., the symmetric case: $B_{1}=B_{2}=B$.

Since the time intervals between the entry of the $P$ particles at $i=L_{1}$ 
is quite long, the number of $P$ particles encountered co-directionally 
by $R_{\ell}$ and that head-on by $R_{r}$  would be almost identical under 
the conditions (a)-(c) above, the most-probable location for the head-on 
meet of the two oppositely moving $R$ particles is expected to be the 
midpoint of the segment $2$ (i.e. at $ i, i+1 \approx L/2, L/2\pm 1 $), and \\
(d) In order to simplify the analytical expressions, we also assume that the 
length of the segment $2$ is $ \approx L/2 $.

Let us define 
\begin{eqnarray}
\frac{\alpha_{q}}{q} \rightarrow \alpha~ , ~\frac{\beta_{q}}{q} 
\rightarrow \beta.
\label{trans}
\end{eqnarray}
as the rescaled initiation and termination rates, respectively, of a
$P$ particle.
Using the well known results for the flux and density profile of TASEP
under open boundary conditions \cite{derrida93,schuetzdomany}, we get
expressions for $J_2$ and $v_2$, in all three possible phases:  In the
low density (LD) phase,
\begin{eqnarray}
J_P=q\alpha(1-\alpha)~~ ,~~ v_P=q(1-\alpha),
\label{ld_1}
\end{eqnarray}
in the high density (HD) phase,
\begin{eqnarray}
J_P=q\beta(1-\beta)~~ ,~~ v_P=q(1-\beta) ,
\label{hd_1}
\end{eqnarray}
and in the maximal current (MC) phase,
\begin{eqnarray}
J_P=\dfrac{q}{4}~~ ,~~ v_P=\dfrac{q}{2} .
\label{mc_1}
\end{eqnarray}
Next, we define the effective velocity of a $R$ particle inside
segment $2$. For $R_{\ell}$
\begin{eqnarray}
v_{R_{\ell}} &=&\begin{cases} B_1(1-\rho) & \mbox{if no P particle in front}\\ 
p_{\text{co}}~(1-\rho) & \mbox{if P particle is in front}  \end{cases} 
\label{eff_vel_2a}
\end{eqnarray}
whereas for $R_{r}$
\begin{eqnarray}
v_{R_{r}} &=& \begin{cases} B_2(1-\rho) &\mbox{if no P particle in front} \\ 
p_{\text{contra}}~(1-\rho) & \mbox{if P particle is in front}  \end{cases} 
\label{eff_vel_2b}
\end{eqnarray}

Therefore, the relative velocities $ v_r $ with which a $R$ particle
approaches a leading $P$ particle, are $ v_{R_{\ell}}-v_P $ and
$ v_{R_{r}}+v_P $ for co-directional and contra-directional encounter,
respectively. The average separation $d$ between the $P$ particles,
i.e., distance headway between the successive particles, in the segment
$2$ is
\begin{equation}
d=\dfrac{1}{\rho}.
\label{distance}
\end{equation}


From expressions (\ref{eff_vel_1}), (\ref{eff_vel_2a}) and
(\ref{eff_vel_2b}), the average interaction times $ \tau_{\text{co}} $ and
$ \tau_{\text{contra}} $, between a $R$ particle and a $P$ particle, during
co-directional and contra-directional encounter, are given by,\\
\begin{eqnarray}
&\tau_{\text{co}}&= \frac{d}{v_{R_{\ell}}-v_{P}} \nonumber \\
&=&\frac{1}{2\rho}\Biggl[\frac{1}{B_1 (1 - \rho) - q (1 - \rho)}
+\frac{1}{p_{\text{co}}(1 - \rho) - q (1 - \rho)}\Biggr], \nonumber \\
\label{time_0}
\end{eqnarray} 
and 
\begin{eqnarray}
&\tau_{\text{contra}}&= \frac{d}{v_{R_{r}}+v_{P}} \nonumber \\
&=&\frac{1}{2\rho}\Biggl[\frac{1}{B_2 (1 - \rho) + q (1 - \rho)}
+\frac{1}{p_{\text{co}} (1 - \rho) + q (1 - \rho)}\Biggr]. \nonumber \\
\label{time_1}
\end{eqnarray}
where we have arrived at the expression for $\tau_{\text{co}}$ assuming it to
be an average of the contributions from the two situations mentioned
in (\ref{eff_vel_2a}). Similarly the expression for $\tau_{\text{contra}}$ is
also the average of the two contributions from the alternative cases
mentioned in (\ref{eff_vel_2b}).

Next, we define $N_{\text{co}}$ and $N_{\text{contra}}$, as the total number of
encounters that a $R$ particle can suffer inside the segment $2$.  We
derive approximate expressions for $N_{\text{co}}$ and $N_{\text{contra}}$. 
When the conditions (a)-(d) are satisfied, $N_{\text{co}}$ and $N_{\text{contra}}$
are given by the expressions
\begin{eqnarray}
N_{\text{co}}&=&N_{\text{contra}}\approx\frac{\rho L}{4}.
\label{number_1}
\end{eqnarray} 
where the factor $L/4$ arises from the fact that each of the $R$
particles has to traverse a distance of $L/4$ to reach the middle of
the segment $2$.
From expressions (\ref{time_0}), (\ref{time_1}) and (\ref{number_1}),
we calculate the total hopping time of a $R$ particle inside segment
$ 2 $, i.e.  $ \tau_{int} $, as a product of total number of
interactions and average encounter time. In the steady state,
$ \tau_{int} $ is given by
\begin{eqnarray}
\tau_{int}&=& N_{\text{co}}~ \tau_{\text{co}} 
+ N_{\text{contra}}~ \tau_{\text{contra}}\nonumber \\
&=&\frac{\rho L}{4}~(\tau_{\text{co}}+\tau_{\text{contra}}).
\label{time_2}
\end{eqnarray}
Further, we calculate the total replication time $\tau$
as a summation of replication times inside segment
$1$ and $3$ and replication time $ \tau_{int} $ inside segment $2$,\\
\begin{eqnarray}
\tau&=&\frac{\tau_{int}}{2}+\frac{L}{4 B_1}\nonumber \\
&=&\frac{\rho L}{8}~(\tau_{\text{co}}+\tau_{\text{contra}})+\frac{L}{4 B_1},
\label{time_3}
\end{eqnarray} 
where the extra factor of $1/2$ in the first term on the right hand 
side is needed to average over the two directions of encounter. 
Note that the dependence of $\tau$ on $\alpha_q$ arises in 
(\ref{time_3}) from the use of the result $\rho(\alpha) = \alpha$ 
for the LD phase of $P$ particles where the relation between $\alpha$ 
and $\alpha_q$ is given by (\ref{trans}).

\begin{figure}[h]
    \begin{center}
        \includegraphics[angle=0,width=0.8\columnwidth]{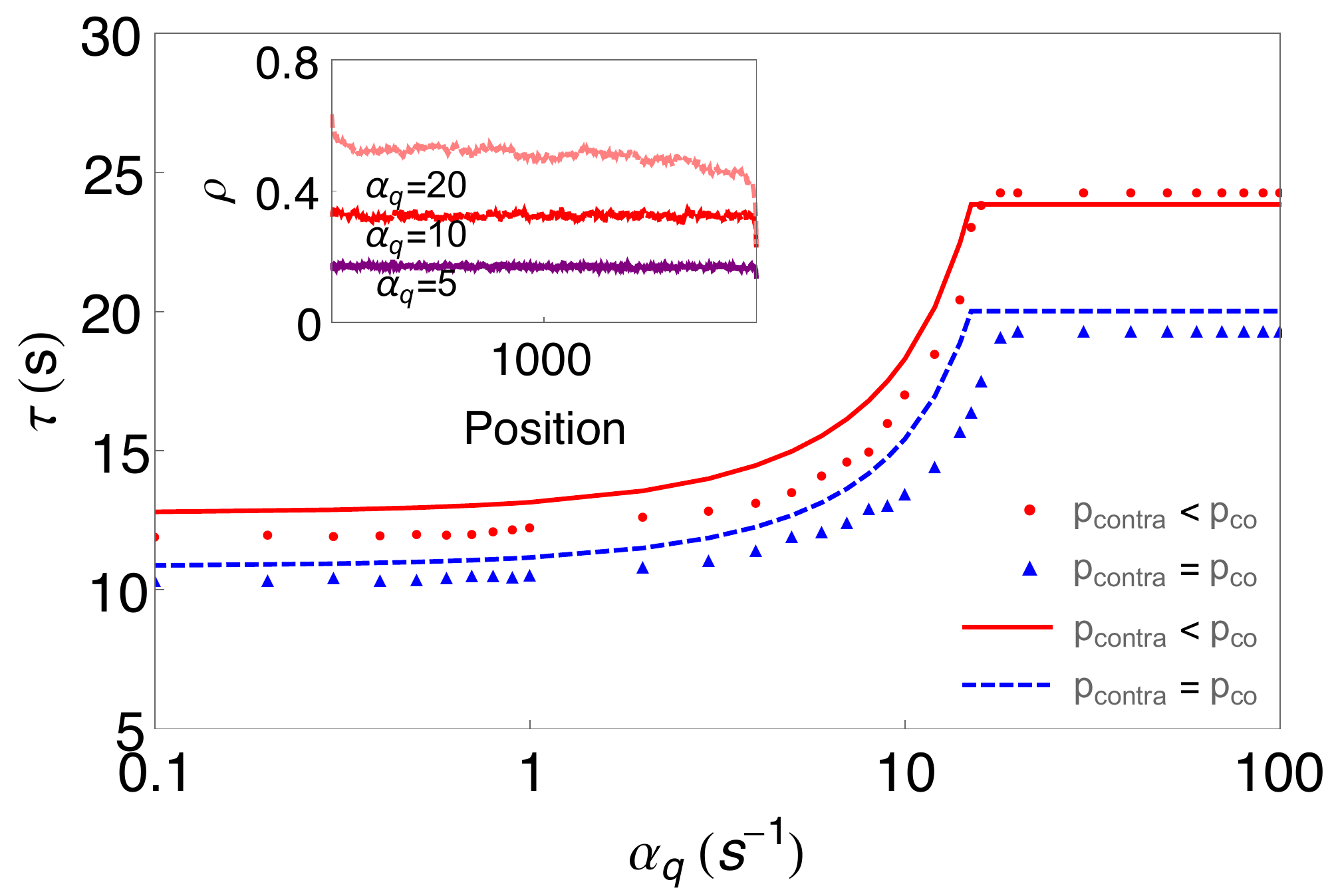}\\[0.02cm]
    \end{center}
    \caption{Variation in average hopping time ($\tau$) is plotted
      with $ \alpha_q $, for two different sets of values of
      $ p_{\text{co}} $ and $ p_{\text{contra}} $. In first case,
      $ p_{\text{co}}=p_{\text{contra}}=30~s^{-1}$ and in second case
      $ p_{\text{co}}=30~s^{-1},p_{\text{contra}}=20~s^{-1} $.  Our theoretical
      predictions, based on heuristic analytical arguments, are drawn
      by continuous curves and numerical data obtained from our
      MC-simulations are shown by discrete points. In inset, we plot
      the density profile of $P$ particles along the lattice, for
      three different values of $ \alpha_q $. The density profile data
      have been obtained only from MC simulations.  The other relevant
      model parameters are
      $ \beta_{q} = 1000 ~s^{-1}, q=30 ~s^{-1}, C=D=0~s^{-1} $. }
    \label{fig-TR_Result}
\end{figure}

In Fig.~\ref{fig-TR_Result}, we show the variation of replication 
time $\tau$  with the rate of transcription initiation (i.e., entry rate 
$ \alpha_q $ of the $P$ particles), for a constant transcription 
termination rate (exit rate of $P$ particles) $\beta_q $. With the 
increase of $\alpha_q$, $ \tau $ increases, and eventually saturates, 
above a critical value of $ \alpha_q $. This behavior is qualitatively 
reproduced by the heuristic analytical arguments. The latter, however, 
tends to slightly overestimate the mean replication time.

The steady state density profiles of the $P$ particle are plotted in
the inset of Fig.~\ref{fig-TR_Result} for few different values of
$\alpha_{q}$. The trend of variation of the profiles with $\alpha_q$ 
is consistent with the transition from the LD phase to MC phase of 
the TASEP of the $P$ particles. For all those values of $ \alpha_q $, 
for which system is in LD phase, particle density $\rho $ increases 
with increase of $ \alpha_q $. Therefore, the total number of
encounters that a $R$ particle can have inside segment $ 2 $ also
increases, which results the increase in $ \tau $.  Above a critical
value of $ \alpha_q $, the TASEP in the segment $ 2 $ makes a transition
to the MC phase where the number density $ \rho $ of the $P$ particles  
and, hence, $ \tau $, becomes independent of $ \alpha_q $.

In Fig.~\ref{fig-TR_Result} we have plotted $\tau$ against
$ \alpha_q $ for two distinct cases. In the first
$p_{\text{co}} = p_{\text{contra}} = 30$ s$^{-1}$ (i.e., the rates of passing is
same irrespective of the direction of encounter). But in the second
case rates of passing are asymmetric, i.e., $p_{\text{contra}} < p_{\text{co}}$,
where $p_{\text{co}} = 30$ s$^{-1}$ and $p_{\text{contra}} = 20$ s$^{-1}$.  The
lower value of $\tau$ in the latter case shows that even if one of the
passing rates decreases, it leads to a lowering of the time needed for
completion of replication because a $R$ particle has to pause for
longer duration.

Next, based on similar heuristic mean-field-type arguments, we derive
analytical expressions for the average replication time in the
opposite limit where $\alpha$ is sufficiently high. In this parameter
regime, at sufficiently small values of $\beta$, the system is in the
high density (HD) phase of TASEP (``termination''-limited regime of
transcription), but makes a transition to the MC phase with the
increase of $\beta$. For the analytical arguments, we assume the same
special scenario (a)-(c) above, i.e., $C = 0 = D$ and $B_1=B_2=B$.

In this case, because of the high value of $\alpha$ the particle
$R_{r}$ is expected to suffer large number of encounters with $P$
particles all of which approach it head-on. Even if it succeeds
entering the segment $2$ through $i=L_{2}$ and move ahead at a slow
pace by passing oncoming $P$ particles, new $P$ particles continue to
make fresh entries into this segment through $i=L_{1}$. Thus, the
number of particles to be bypassed by $R_{r}$ keep increasing as time
passes till $R_{r}$ exits the segment $2$ through $i=L_{1}$.  In
contrast, the particle $R_{\ell}$ encounters far fewer $P$ particles
because, after it enters the segment $2$, the new entrant $P$
particles would be falling behind it and even some of those in front
would make their exit from $i=L_{2}$ before $R_{\ell}$ catches up
co-directionally from behind.  Therefore, we make the simplifying
assumption (perhaps, slight oversimplification) that the particle
$R_{r}$ remains stalled at $i=L_{2}+1$ and replication is completed
only when $R_{\ell}$ reached $i=L_{2}$.

The average number of $P$ particles within the segment from $L_{1}$ to
$L_{2}$ is $(L_{2}-L_{1})\rho$ where $\rho$ is the average number
density of the $R$ particles in this segment. The average spatial gap
between the $P$ particles, as given by eq.~(\ref{distance}), is $1/\rho$
and the number of gaps to be covered by a $R$ particle is
$(L_{2}-L_{1})\rho$. Therefore, the total time spent by the $R$
particle in exchanging its position with the co-directionally moving
$P$ particles is
\begin{equation}
\tau_{\text{exch}} = (L_{2} - L_{1}) \rho/p_{\text{co}}.
\end{equation} 
The effective velocity of a $R$ particle in the segment between
$L_{2}$ and $L_{1}$ is $B-q$.  The total time spent by the $R$
particle in covering all the gaps by forward hopping is
\begin{equation}
\tau_{\text{hop}} = (L_{2}-L_{1})\rho \frac{1}{\rho}\frac{1}{(B-q)} 
= (L_{2}-L_{1})/(B-q). 
\end{equation} 
The time taken by the $R$ particle to reach $L_{1}$ from $i=1$ is 
\begin{equation}
\tau_{arr} = L_{1}/B.
\end{equation}
Thus, finally, the total time taken to complete replication is 
\begin{equation}
\tau = \frac{(L_{2} - L_{1}) \rho}{p_{\text{co}}} 
+ \frac{(L_{2}-L_{1})}{(B-q)} + \frac{L_{1}}{B}\,.
\label{eq-HD2MC}
\end{equation}

\begin{figure}[h]
    \begin{center}
        \includegraphics[angle=0,width=0.9\columnwidth]{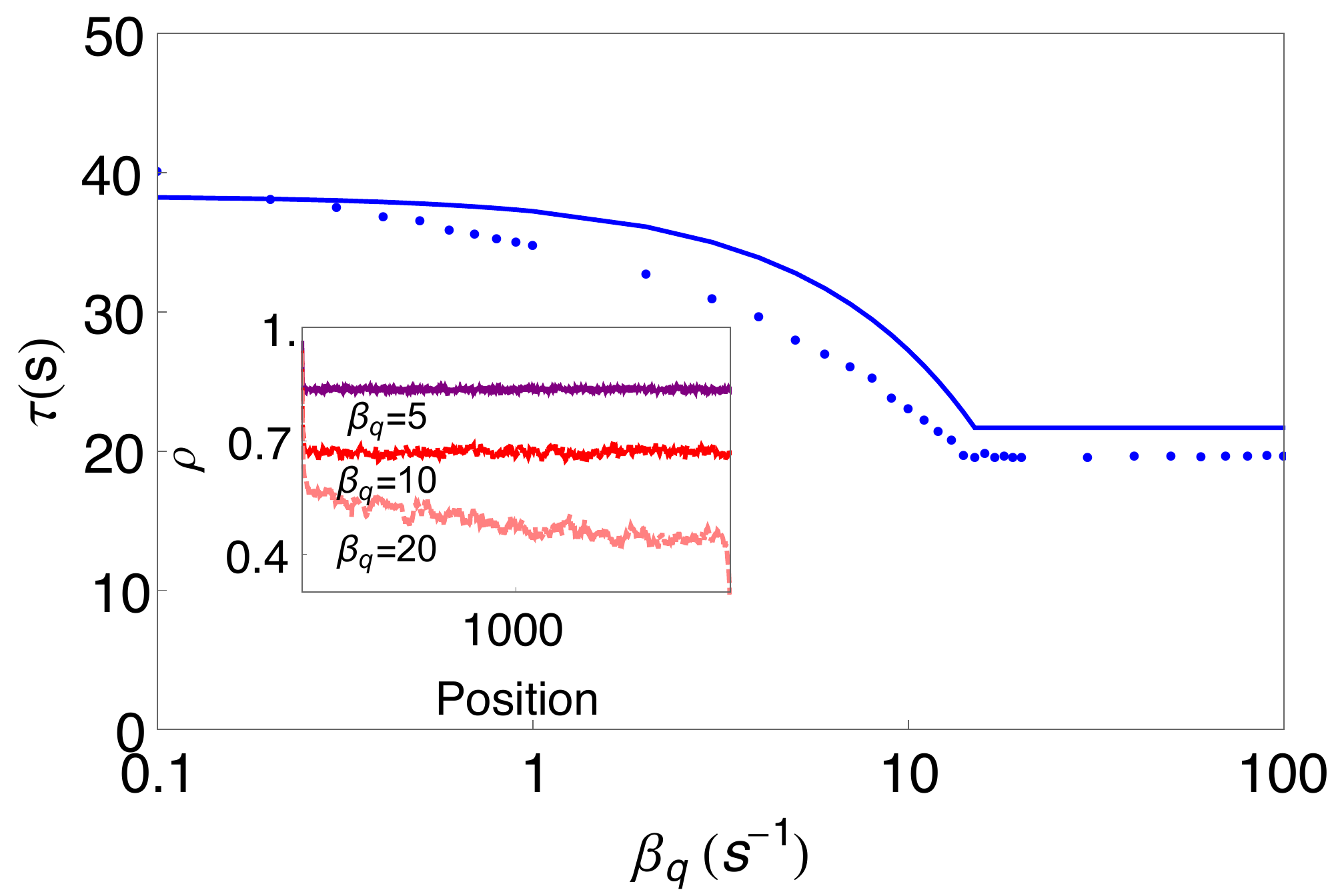}\\[0.02cm]
    \end{center}
    \caption{ The replication time $\tau$ is plotted as a function
      of $ \beta_q $.  Our theoretical predictions, based on heuristic
      analytical arguments, are drawn by continuous curves and
      numerical data obtained from our MC-simulations are shown by
      discrete points. In inset, we plot the density profile of $P$
      particles along the lattice, for three different values of
      $ \beta_q $. The density profile data have been obtained only
      from MC simulations.  The other relevant model parameters are
      $ \alpha_{q}=100 s^{-1}, q=30 s^{-1},
      p_{\text{co}}=p_{\text{contra}}=30 s^{-1}$.}
    \label{fig-TR_Result_opp}
\end{figure}

In Fig.~\ref{fig-TR_Result_opp}, we show the variation of replication
time $\tau$ with the rate of transcription termination (i.e., exit
rate $\beta_q $ of the $P$ particles), for a constant transcription
initiation rate (entry rate of $P$ particles) $\alpha_q $.  Note that
the dependence of $\tau$ on $\alpha_q$ arises in (\ref{eq-HD2MC}) from
the use of the result $\rho(\beta) = 1 - \beta$ for the HD phase of
$P$ particles where the relation between $\beta$ and $\beta_q$ is
given by (\ref{trans}).  With the increase of $\beta_q$, $ \tau $
decreases, and eventually saturates, above a critical value of
$ \beta_q $.  This behavior is qualitatively reproduced by the
heuristic analytical arguments which slightly 
overestimate the mean replication time.
The steady state density profiles of the $P$ particle
are plotted in the inset of Fig.~\ref{fig-TR_Result_opp} for few
different values of $\beta_{q}$. The trend of variation of the
profiles with $\beta_q$ is consistent with the transition from the HD
phase to MC phase of the TASEP of the $P$ particles. For all those
values of $ \beta_q $, for which system is in HD phase, particle
density $\rho $ decreases with increase of $ \beta_q $. Therefore, the
total number of encounters that a $R$ particle can have inside segment
$ 2 $ also decreases, which results the decrease in $ \tau $.  Above a
critical value of $ \beta_q $, the TASEP in the segment $ 2 $ makes a
transition to the MC phase where the number density $ \rho $ of the
$P$ particles and, hence, $ \tau $, becomes independent of
$ \beta_q $.

In the absence of collapse of the replication fork ($C=0$) and premature 
detachment of RNAP ($D=0$), the replication time is essentially decided 
by the density of the RNAP motors (i.e., $R$ particles). Since one or two 
$R$ particles make hardly any noticeable perturbation of the density that  
is prescribed by the exact theory for a pure TASEP of $P$ particles, the 
expressions (\ref{time_3}) and (\ref{eq-HD2MC}) for the replication time 
$\tau$ are in excellent agreement with the corresponding data obtained 
from MC simulation of the model.

\subsection{Histograms of the Number of Successful and 
Unsuccessful Replication Events}


\begin{figure}[t]
\begin{center} 
(a) \\
\includegraphics[angle=0,width=1.0\columnwidth]{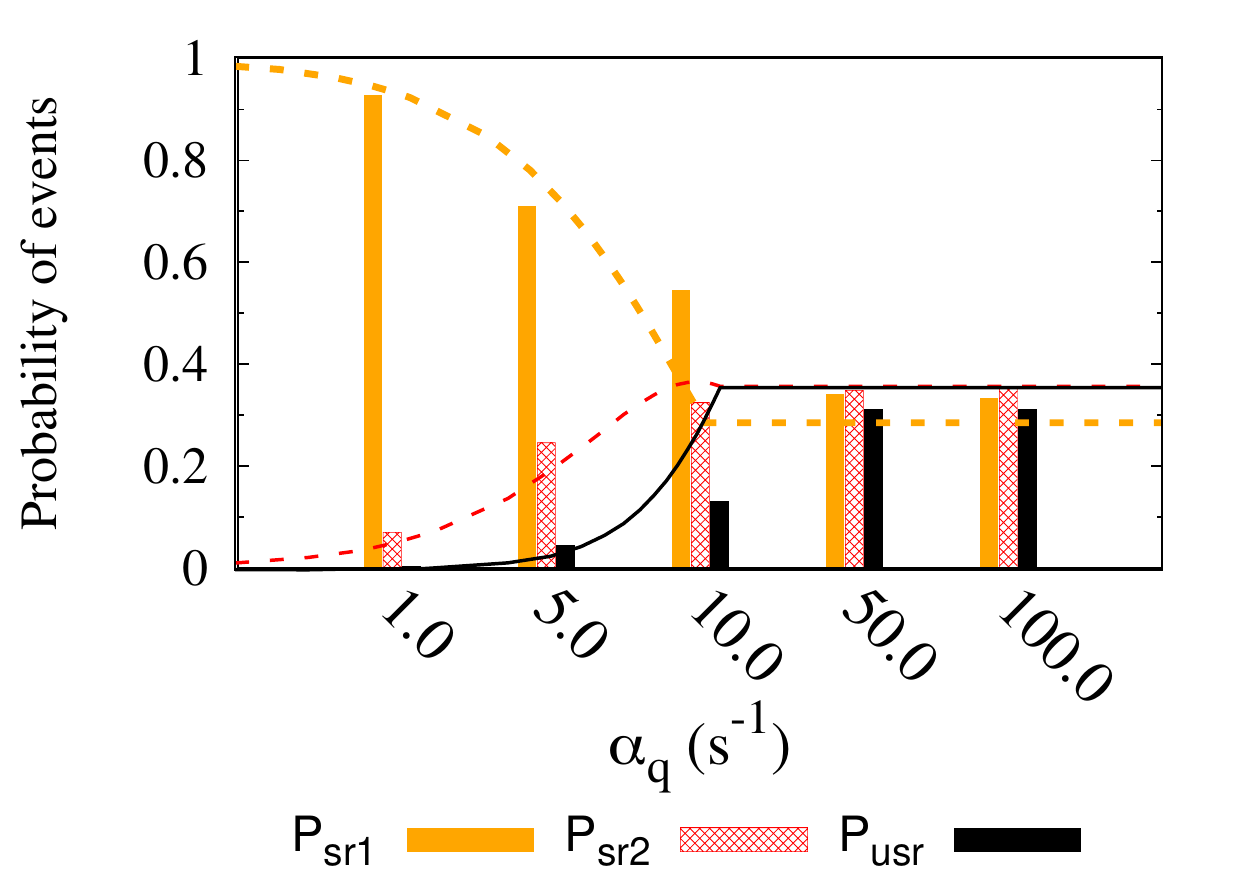}\\[0.02 cm]
(b) \\
\includegraphics[angle=0,width=1.0\columnwidth]{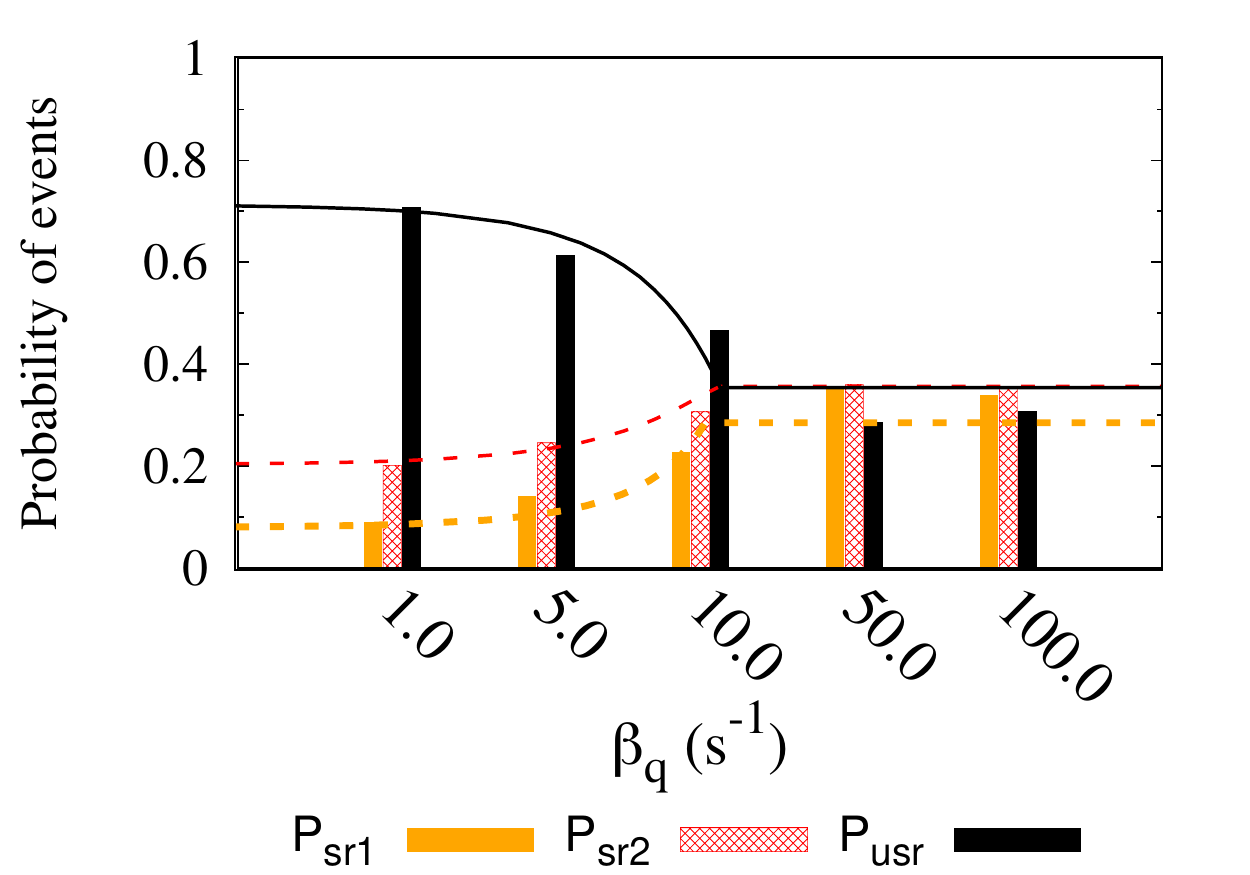}\\[0.02 cm]
\end{center}
\caption{Distribution of sr1, sr2 and usr in the special case
  $C \neq 0$, $D=0$ is plotted for five different values of
  (a) $\alpha_{q}$, for fixed $\beta_{q}=1000~s^{-1}$ and (b)
  $\beta_{q}$, for fixed $\alpha_{q}=1000~s^{-1}$ . The data used for
  the bar plots have been obtained by MC-simulations. Lines have been
  obtained from the analytical expressions (\ref{eq-sr11}),
  (\ref{eq-sr21}) and (\ref{eq-usr1}). Dotted line corresponds to
  $P_{\text{sr1}}$, dashed line corresponds to $P_{\text{sr2}}$ and continuous line
  corresponds to $P_{\text{usr}}$. The other relevant parameters used in this
  figure are $L=2000$, $q=30~s^{-1}$, $C = 0.05~s^{-1}$, $D = 0$ and
  $p_{\text{co}} = p_{\text{contra}} = p = 20~s^{-1}$.}
	\label{fig-hist}
\end{figure}

\begin{figure}[h]
    \begin{center} 
(a)\\
        \includegraphics[angle=0,width=1.0\columnwidth]{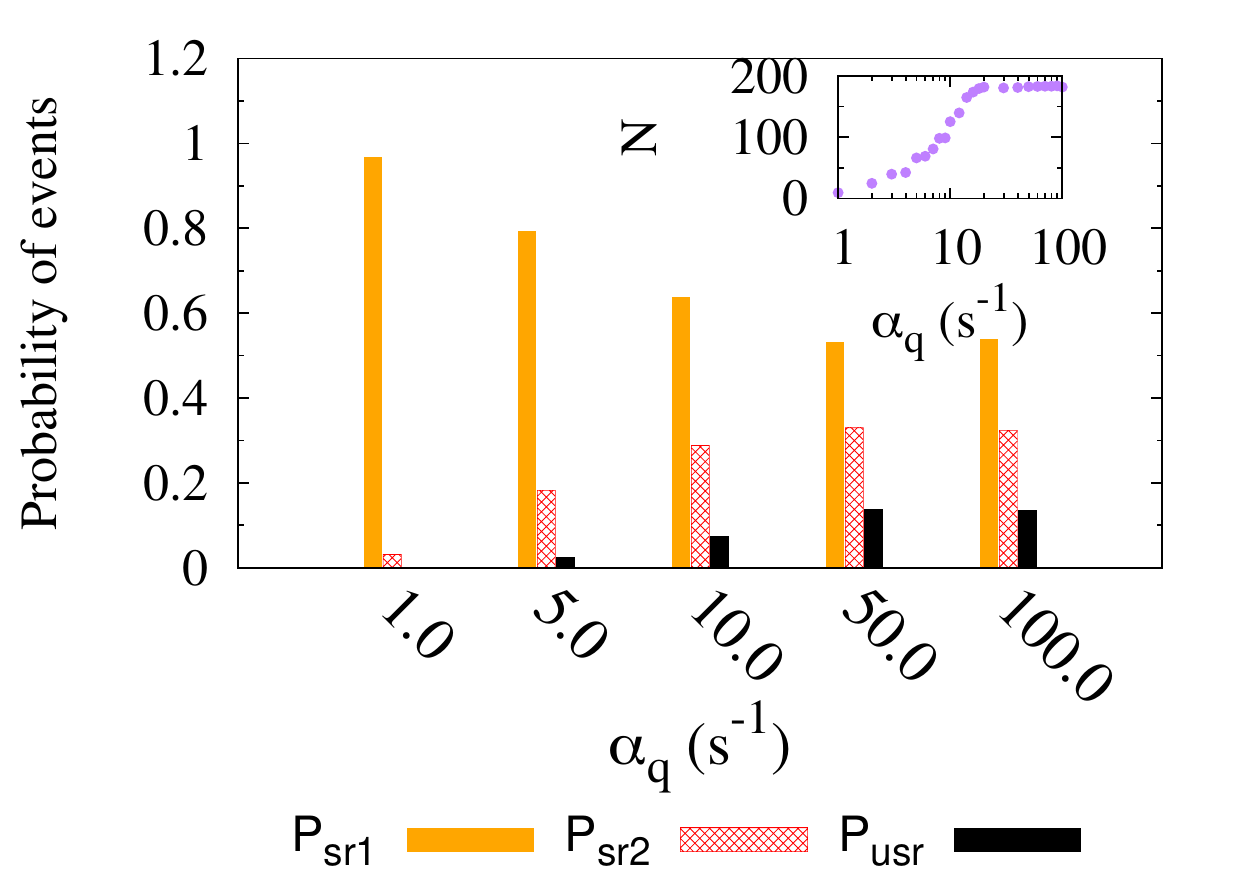}\\[0.02cm]
(b)\\
        \includegraphics[angle=0,width=1.0\columnwidth]{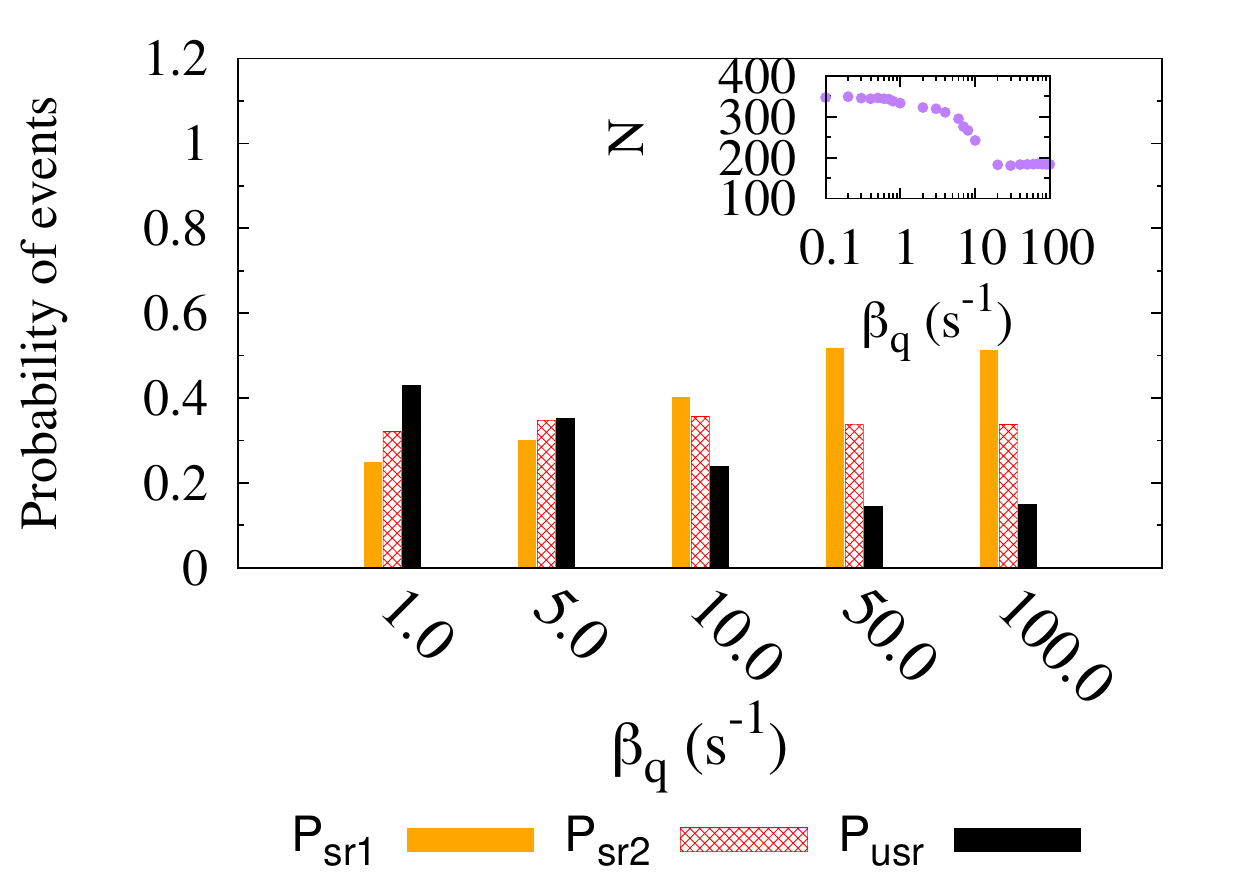}\\[0.02cm]
    \end{center}
    \caption{Distribution of sr1, sr2 and usr in the general case
      $C\neq 0$, $D \neq 0$ is plotted for five different values of
      (a) $\alpha_{q}$, for fixed $\beta_{q}=1000~s^{-1}$ and (b)
      $\beta_{q}$, for fixed $\alpha_{q}=1000~s^{-1}$.  In inset we
      plot the variation in $ N $ (i.e. average number of $P$ particle
      detached from the track during their encounter with $R$
      particles) with (a) $ \alpha_q$ and (b) $\beta_{q}$.  These data
      have been obtained only from MC-simulations. The other relevant
      parameters used in this figure are $L=2000$, $q=30~s^{-1}$,
      $C = 0.05~s^{-1}$, $D = 10~s^{-1}$ and
      $p_{\text{co}} = p_{\text{contra}} = p = 20~s^{-1}$.}
    \label{fig-Histogram}
\end{figure}

In this subsection we show the effects of transcription on
replication. For this purpose, we calculate how the distributions of
the three processes, namely, sr1, sr2 and usr are affected by the
encounter of $R$ particles with the $P$ particles.\\

\noindent $\bullet$ {\bf Special case of $C \neq 0$ and $D = 0$}\\

Nonzero $C$ gives rise to two other alternative scenarios. If only one
of the $R$ particles collapses and the other does not replication is
completed via the alternative route that we defined as sr2. Similarly,
collapse of both the $R$ particles leads to nonzero probability of
usr. Note that any increase in the probabilities of sr2 or usr, or
both, cause reduction in the probability of sr1 because
$P_{\text{sr1}}+P_{\text{sr2}}+P_{\text{usr}}=1$.

Suppose, on the average, the total number of $P$ particles in the
segment between $i=L_{1}$ and $i=L_{2}$ is $N$. If $N$ events of
passing in the segment between $i=L_{1}$ and $i=L_{2}$ is required for
completion of replication without suffering collapse of either of the
two replication forks, then the probability of sr1 is
\begin{eqnarray}
P_{\text{sr1}} &=& \biggl(\frac{p}{p+C}\biggr)^{N} 
\label{eq-sr11}
\end{eqnarray}
For sr2, one of the forks ($R$ particles) has to collapse while the
remaining stretch of the segment between $i=L_{1}$ to $i=L_{2}$ is
covered by the surviving fork.  One of the forks may collapse after
passing $n$ number of $P$ particles; the probability of its occurrence
is $[C/(p+C)][p/(p+C)]^{n}$; the probabilty that the surviving fork
passes the other remaining $P$ particles is $[p/(p+C)]^{N-n}$.  Thus,
the probability of sr2 is
\begin{eqnarray}
P_{\text{sr2}}&=&    \sum_{n=0}^{N-1}    \biggl[ \biggl(\frac{C}{p+C}\biggr) 
\biggl(\frac{p}{p+C}\biggr)^{n}\biggr]
           \biggl(\frac{p}{p+C}\biggr)^{N-n} 
\nonumber \\
&=&  N \biggl(\frac{C}{p+C}\biggr) \biggl(\frac{p}{p+C}\biggr)^{N} 
\label{eq-sr21}
\end{eqnarray}
Exploiting normalization, we get the probability for usr
\begin{eqnarray}
P_{\text{usr}}&=& 1-P_{\text{sr1}}-P_{\text{sr2}}.
\label{eq-usr1}
\end{eqnarray}
Note that $N=\rho (L_{2}-L_{1})$ is the average number of $P$
particles in the interaction segment between $i=L_1$ and $i=L_2$. 
In the LD regime of $P$ particles $\rho=\alpha=\alpha_{q}/q$.\\

In Fig.~\ref{fig-hist}, we plot the distributions of 'sr1', 'sr2' and
`usr' as histograms for (a) five distinct values of $ \alpha_q $ and a
constant value of $ \beta_q $, and (b) five distinct values of
$ \beta_q $ and a constant value of $ \alpha_q $. 
The analytic approximations (\ref{eq-sr11})--(\ref{eq-usr1})
  reproduce qualitatively the behavior observed in the MC simulations. 
For a given sufficiently high value of $\beta_q$, segment $ 2 $ is in
the LD phase at small values of $ \alpha_q $. In this regime the
number of eventual collapse of a $R$ particle during its encounters
with $P$ particles is negligibly small. Therefore, for these small
values of $ \alpha_q $, number of events of the type 'sr2' and `usr'
are low and, hence the probability of sr1 is very weakly affected (see
Fig.~\ref{fig-hist}(a)). As $ \alpha_q $ increases further, the number
of eventual collapse increases because of the increasing number of
encounters with $P$ particles which is reflected in the significant
increase in 'sr2' and `usr' in Fig.~\ref{fig-hist}(a). Increase in the
probabilities of sr2 and usr results in the corresponding decrease in
the probability of sr1 because of the normalization of the
probabilities mentioned above. Number of the events 'sr1', 'sr2' and
`usr' attain their respective saturation values as $ \alpha_q $
increases above the critical value where the transition from LD phase
to MC phase takes place (see Fig.~\ref{fig-hist}(a)).

Similarly, for a sufficiently high value of $\alpha_q$, with
increasing $\beta_q$ the $P$ particles exhibit a transition from 
the HD phase to the MC phase. Consequently, the decrease in the 
frequency of encounter of the $P$ particles with the $R$ particles. 
The likelihood of collapse of both the $R$ particles in any run 
decreases as indicated by the increase of the probability of sr2. 
The concomitant increase of the probabiity of sr is also shown 
in Fig.~\ref{fig-hist}(b).\\

\noindent $\bullet$ {\bf Special case of $C \neq 0$ and $D \neq 0$}\\

Now, we consider the general case of our model allowing for the
possibilities that $C \neq 0$ and $D \neq 0$. As we have already done 
in the restricted case of $C \neq 0$, $D=0$, 
we characterize the effect of nonzero $C$ and $D$ also 
in terms of the statistics of 'sr1', 'sr2' and 'usr'. 

In Fig.~\ref{fig-Histogram}(a) and (b), we  plot the distributions of the 
events 'sr1', 'sr2' and 'usr' as histograms for (a) five different values of
$ \alpha_q $ at a constant high value of $ \beta_q $ and (b) five different 
values of$ \beta_q $ at a constant high value of $ \alpha_q $. The trends of 
variation of these three probabilities are explained by the transition to the 
MC phase from (a) LD phase and (b) HD phase.

In the inset of Fig.~\ref{fig-Histogram}, we display the effects of 
replication on transcription. We show the variation in the number 
$N $ of $P$ particles that detach from the lattice when they encounter 
a $R$ particle, for a given rate $ \alpha_q $. The
trend of variation and the physical reason for this trend is also well
explained by the transition from LD phase to MC phase.

\subsection{Distribution of Detachments of $P$ Particles}

In MC simulations, we measure the time intervals between two consecutive
$ P $ detachment events as $ \delta t_1 $,
$ \delta t_2 \dots \delta t_n $, if $ n+1 $ detachments takes place in
a single MC simulation run. Since this is a stochastic process these
time intervals $ \delta t_1 $, $ \delta t_2 \dots \delta t_n $ are, in
general, different from each other. We compute the number of
consecutive $ P $ detachment events corresponding to a given interval
$ \delta t $, i.e. if two time intervals are identical
($ \delta t_i=\delta t_j= \delta t$), then, number of consecutive
$ P $ detachment events with time interval $ \delta t $ is $ 2 $. We
repeat the procedure over $ 10000 $ MC simulation runs to calculate
total number of consecutive $ P $ detachment events with given
interval $ \delta t $, and then we divide this number with number of
MC simulation runs, i.e. $ 10000 $, to calculate the average number
$ N_{\alpha} $ of consecutive $P$ detachments within the time interval 
$ \delta t $ for a fixed rate $ \alpha_q$.
\begin{figure}[h]
    \begin{center}
        \includegraphics[angle=0,width=0.8\columnwidth]{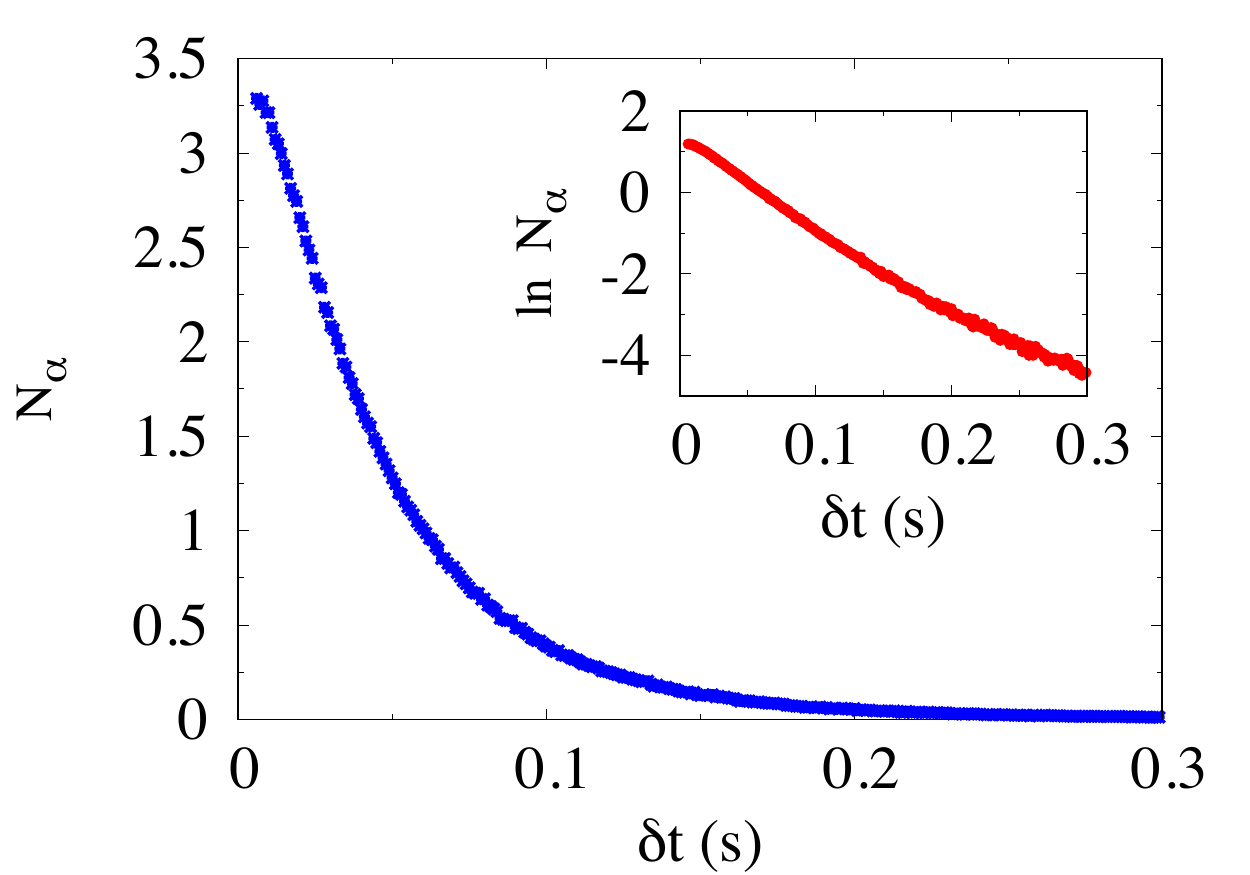}\\[0.02cm]
    \end{center}
    \caption{ Distribution of $ N_{\alpha} $ is plotted with
      $ \delta t$ for a constant $ \alpha_q =100~s^{-1}$. In inset we
      plot $ N_{\alpha} $ with $ \delta t $ on a semi-log axis to show
      the exponential fall of $ N_{\alpha} $ with $ \delta t $
      . These data have been obtained only by MC-simulations. The other
      relevant parameters used in this figure are $C=0.05~s^{-1}$,
      $D=10~s^{-1}$, $p_{\text{co}}=20~s^{-1}$ and $p_{\text{contra}}=20~s^{-1}$.}
    \label{fig-Distribution}
\end{figure}

In Fig.~\ref{fig-Distribution} we show the variation in
$ N_{\alpha} $ with $ \delta t $ and we find that $ N_{\alpha} $ falls
exponentially as the time interval between two consecutive $ P $
detachment events increases. To confirm the exponential behavior, in 
the inset we show the variation in $ N_{\alpha} $ on a semi-log axis with
$ \delta t $.


\section{Summary and conclusion} 

In this paper we have developed the first minimal model that captures
the key kinetic rules for the resolution of conflict between
transcription and concomitant replication of the same stretch of
DNA. This model has been formulated in terms of a two-species
exclusion process where one species of particles (denoted by $P$)
represents the RNA polymerase motors while the two members of the
other species (denoted by $R$) represent the two replication forks. 

In contrast to all the multi-species exclusion models reported so far,
the allowed populations of the two species are quite different in our
model. A maximum of only two $R$ particles are allowed to enter the
lattice; imposition of this restriction on the number of $R$ particles
is motivated by the fact that none of the segments of DNA should be
replicated more than once during the life time of a cell. In sharp
contrast, the number of $P$ particles is not restricted except for the
control of their population by the rate constants for their entry,
exit and hopping.  This choice is consistent with the fact that the
multiple rounds of transcription of the same segment of DNA is not
only possible but resulting synthesis of multiple identical
transcripts is also desirable for the proper biological function of
the cell.  Moreover, all the $P$ particles move co-directionally, from
left to right whereas one of the $R$ particles (namely, $R_{\ell}$)
moves from left to right while the other $R$ particle (namely $R_{r}$)
approaches it head-on from the opposite end.  Another distinct feature
of this model is that the lattice consists of three segments; the
encounters of RNAP motors ($P$ particles) with the replication fork
($R$ particles) are confined within the middle segment (segment $2$)
whereas only the $R$ particles can occupy the sites in the segments
$1$ and $3$.

By a combination of analytical treatment, based on heuristic arguments, 
and Monte Carlo simulations we have analyzed the effects of the RNA 
polymerase motor traffic on the DNA replication and vice versa. More
specifically, we have shown how the transition from the Low-density
phase to the Maximal Current phase and that from the high-density 
phase to the maximal Current phase of $P$ traffic affects the not only
the total time required for successful completion of replication but
also how the statistics of the successful and unsuccessful replication
events are also affected.

Any attempt of direct comparison between the theoretical predictions
of our model with the experimental data may be premature at this
stage. There are some important features of DNA replication in
eukaryotic cells that we hope to incorporate in future extensions of
our model. For example, even after two replication forks begin
approaching each other head-on, new pairs of replication forks can
nucleate in the unreplicated segment of the DNA in between the two.
However, the price to be paid for more realistic and more detailed
would be to sacrifice the possibility of analytical treatments even on
the basis of heuristic arguments. Nevertheless, computer simulations
would still provide some mechanistic insight into the causes and
consequences of the transcription-replication conflict.

\vspace{0.7cm}

This work has been supported by J.C. Bose National Fellowship (DC),
``Prof. S. Sampath Chair'' Professorship (DC), by UGC Senior
Research Fellowship (BM) and the German Science Foundation (DFG)
under grant SCHA 636/8-2 (AS).



\begin{thebibliography}{99}

\bibitem{derrida98} B. Derrida: 
 Phys. Rep. \textbf{301}, 65 (1998)

\bibitem{schutz00} G.M. Sch\"utz: {\it Integrable stochastic many-body
    systems}, in: Phase Transitions and Critical Phenomena
  \textbf{19}, 1 (2000), Ed.  C. Domb and J L Lebowitz (London:
  Academic)

\bibitem{Schadschneider10} A. Schadschneider, D. Chowdhury,
  K. Nishinari:  {\it Stochastic Transport in Complex Systems: From
  Molecules to Vehicles} (Elsevier, 2010)

\bibitem{mallick15} K. Mallick, Physica A {\bf 418}, 17-48 (2015).

\bibitem{MacDonald68} C.T. MacDonald, J.H. Gibbs, A.C. Pipkin:
Biopolymers \textbf{6}, 1 (1968).

\bibitem{macdonald69} C. MacDonald and J. Gibbs, Biopolymers,
{\bf 7}, 707 (1969). 

\bibitem{basu07} A. Basu and D. Chowdhury, Phys. Rev. E {\bf 75}, 021902 (2007). 

\bibitem{gccr09} A. Garai, D. Chowdhury,  D. Chowdhury and T.V. Ramakrishnan,  Phys. Rev. E {\bf 80}, 011908 (2009). 

\bibitem{zia11} R.K.P. Zia, J.J. Dong and B. Schmittmann, J. Stat. Phys. {\bf 144}, 405 (2011). 

\bibitem{lin11} C. Lin, G. Steinberg and P. Ashwin, J. Stat. Mech. Theor. Expt. P09027 (2011).

\bibitem{greulich12} P. Greulich, L. Ciandrini, R.J. Allen and M.C. Romano,  Phys. Rev. E {\bf 85}, 011142 (2012).

\bibitem{chowdhury08} D. Chowdhury, A. Garai and J.S. Wang, Phys. Rev. E  {\bf 77}, 050902(R) (2008).

\bibitem{oriola15} D. Oriola, S. Roth, M. Dogterom and J. Casademunt, Nat. Commun. {\bf 6}, 8025 (2015).

\bibitem{sugden07} K.E.P. Sugden, M.R. Evans, W.C.K. Poon and N.D. Read,  Phys. Rev. E {\bf 75}, 031909 (2007).

\bibitem{evans11} M.R. Evans, Y. Kafri, K.E.P. Sugden and J. Tailleur,  J. Stat. Mech. Theor. Expt., P06009 (2011).

\bibitem{chai09} Y. Chai, S. Klumpp, M.J.I. M\"uller and R. Lipowsky,  Phys. Rev. E {\bf 80}, 041928 (2009). 

\bibitem{ebbinghaus09} M. Ebbinghaus and L. Santen, J. Stat. Mech.: Theor. Expt. P03030 (2009). 

\bibitem{ebbinghaus10} M. Ebbinghaus, C. Appert-Rolland and L. Santen, Phys. Rev. E {\bf 82}, 040901 (R) (2010). 

\bibitem{muhuri10} S. Muhuri and I. Pagonabarraga, Phys. Rev. E {\bf 82}, 021925 (2010).

\bibitem{neri11} I. Neri, N. Kern and A. Parmeggiani, Phys. Rev. Lett. {\bf 107}, 068702 (2011).

\bibitem{neri13a} I. Neri, N. Kern and A. Parmeggiani, Phys. Rev. Lett. {\bf 110}, 098102 (2013).

\bibitem{neri13b} I. Neri, N. Kern and A. Parmeggiani, New J. Phys. {\bf 15},  085005 (2013).

\bibitem{curatolo16} A.I. Curatolo, M.R. Evans, Y. Kafri and J. Tailleur, J. Phys. A: Math. Theor.  {\bf 49}, 095601 (2016).

\bibitem{klein16} S. Klein, C. Appert-Rolland and M.R. Evans, J. Stat Mech. 093206 (2016).

\bibitem{parmeggiani04} A. Parmeggiani, T. Franosch and E. Frey, Phys. Rev. E {\bf 70}, 046101 (2004).

\bibitem{graf17} I.R. Graf and E. Frey, Phys. Rev. Lett. {\bf 118}, 128101 (2017).

\bibitem{chowdhury05} D. Chowdhury, A. Schadschneider and K. Nishinari:
Phys. Life Rev. {\bf 2}, 318 (2005). 

\bibitem{chou11} T. Chou, K. Mallick and R.K.P. Zia: 
Rep. Prog. Phys. {\bf 74}, 116601 (2011).

\bibitem{chowdhury13} D. Chowdhury:  
Phys. Rep. \textbf{529}, 1-197 (2013).

\bibitem{rolland15} C. Appert-Rolland, M. Ebbinghaus and L. Santen:
Phys. Rep. {\bf 593}, 1 (2015).

\bibitem{kolomeisky15} A.B. Kolomeisky: 
{\it Motor proteins and molecular motors}, (CRC Press, 2015). 

\bibitem{kornbergbook} A. Kornberg and T. Baker: 
{\it DNA Replication} (University Science Books, 1992). 

\bibitem{tripathi08} T. Tripathi and D. Chowdhury: 
Phys. Rev. E {\bf 77}, 011921 (2008). 

\bibitem{klumpp08} S. Klumpp and T. Hwa: 
Proc. Natl. Acad. Sci. {\bf 105}, 18159 (2008).

\bibitem{klumpp11} S. Klumpp: J. Stat. Phys. {\bf 142}, 1252 (2011).

\bibitem{sahoo11} M. Sahoo and S. Klumpp: EPL {\bf 96}, 60004 (2011). 

\bibitem{ohta11} Y. Ohta, T. Kodama and S. Ihara: 
Phys. Rev. E {\bf 84}, 041922 (2011). 

\bibitem{wang14} J. Wang, B. Pfeuty, Q. Thommen, M.C. Romano and
  M. Lefranc: Phys. Rev. E {\bf 90}, 050701(R) (2014).

\bibitem{jun05a} S. Jun, H. Zhang and J. Bechhoefer:
Phys. Rev. E {\bf 71}, 011908 (2005). 

\bibitem{jun05b} S. Jun and J. Bechhoefer:
Phys. Rev. E {\bf 71}, 011909 (2005). 

\bibitem{bechhoefer07} J. Bechhoefer and B. Marshall: 
Phys. Rev. Lett. {\bf 98}, 098105 (2007). 

\bibitem{yang08} S.C. Yang and J. Bechhoefer:
Phys. Rev. E {\bf 78}, 041917 (2008). 

\bibitem{baker12} A. baker, B. Audit, S.C.H. Yang, J. Bechhoefer and 
A. Arneodo: Phys. Rev. Lett. {\bf 108}, 268101 (2012). 

\bibitem{retkute11} R. Retkute, C. A. Nieduszynski, A. de Moura: 
Phys. Rev. Lett. {\bf 107}, 068103 (2011). 

\bibitem{retkute12} R. Retkute, C. A. Nieduszynski, A. de Moura: 
Phys. Rev. E \textbf{86}(3), 031916 (2012). 

\bibitem{yang09} S.C. Yang, M.G. Gauthier and J. Bechhoefer: in {\it Methods in  Mol. Biol, DNA Replication, vol. 521}, eds. S. Vengrova and J.Z. Dalgaard, (Springer 2009). 

\bibitem{hyrien10} O. Hyrien and A. Goldar: Chromosome Research {\bf 18}, 147 (2010).  

\bibitem{kim12} N. Kim and S. Jinks-Robertson, Nat. Rev. Genet. {\bf 13}, 204 (2012).

\bibitem{merrikh12} H. Merrikh, Y. Zhang, A.D. Grossman and J.D. Wang: Nat. Rev. Microbiol. {\bf 10}, 449 (2012). 

\bibitem{helmrich13} A. Helmrich, M. Ballarino, E. Nudler and L. Tora: Nat. Struct. Mol. Biol. {\bf 20}, 412 (2013). 

\bibitem{muse16} T. Garcia-Muse and A. Aguilera: Nat. Rev. Mol. Cell Biol. {\bf 17}, 553 (2016). 

\bibitem{pomerantz10} R. T. Pomerantz, M. O'Donnell:
Cell Cycle (Georgetown, Tex.) \textbf{9}(13), 2537-2543 (2010).

\bibitem{parmeggiani03}  A. Parmeggiani, T. Franosch, E. Frey: 
Phys. Rev. Lett. \textbf{90}, 086601 (2003).

\bibitem{chowdhury00} D. Chowdhury, L. Santen, A. Schadschneider: 
Phys. Rep. \textbf{329}, 199 (2000).

\bibitem{derrida93} B. Derrida, M.R. Evans, V. Hakim, V. Pasquier:
J. Phys. A: Math. Gen. \textbf{26}, 1493 (1993)

\bibitem{schuetzdomany} G.M. Sch\"utz, E. Domany: 
J. Stat. Phys. \textbf{72}, 277 (1993)

\bibitem{Liu} B. Liu, B.M. Alberts: 
Science \textbf{267}, 1131-1137 (1995)

\bibitem{Soultanas} P. Soultanas, Transcription {\bf 2:3}, 140 (2011).

\bibitem{French} S. French: 
Science \textbf{258}, 1362 (1992)

\bibitem{Mirkin} E.V. Mirkin and S.M. Mirkin: 
Microbiology and Molecular Biology Reviews \textbf{71}, 13-35 (2007)

\bibitem{Rothstein} R. Rothstein, B. Michel, S. Gangloff: 
Genes Development \textbf{14}, 110 (2000)

\end{thebibliography}
\end{document}